\documentclass[symmetry,article,accept,pdftex,moreauthors]{Definitions/mdpi} 
\newcommand{\apj}{Astrophys. J.}

\newcommand{\mnras}{MNRAS}

\firstpage{1} 
\pubvolume{1}
\issuenum{1}
\articlenumber{0}
\pubyear{2024}
\copyrightyear{2024}
\externaleditor{Academic Editor: Manuel Gadella}
\datereceived{19 June 2024} 
\daterevised{2 July 2024} 
\dateaccepted{6 July 2024} 
\datepublished{ } 
\hreflink{https://doi.org/} 

\Title{Multidimensional Representation of Semantic Relations between Physical Theories, Fundamental Constants and Units of Measurement with Formal Concept Analysis
}

\TitleCitation{Multidimensional Representation of Semantic Relations between Physical Theories, Fundamental Constants and Units of Measurement with Formal Concept Analysis
}

\Author{{Mariana Espinosa-Aldama} 
 $^{1,*,\dagger}$\orcidA{} and 
  Sergio Mendoza $^{2,*,\dagger}$\orcidB{} }

\AuthorNames{Mariana Espinosa-Aldama and Sergio Mendoza}

\AuthorCitation{Espinosa-Aldama, M.; Mendoza, S.}

\address{%
$^{1}$ \quad {Instituto} 
 de Investigaciones en Matem\'aticas Aplicadas y en Sistemas, {Universidad} Nacional Aut\'onoma de M\'exico, 
Ciudad Universitaria {04510}, 
 Ciudad de M\'exico,
M\'exico\\
$^{2}$ \quad {Instituto} de Astronom\'{\i}a, {Universidad} Nacional
Aut\'onoma de M\'exico, {AP~70-263}, Ciudad Universitaria 04510, 
 Ciudad de M\'exico,
M\'exico
}

\corres{Correspondence: mariana.espinosa@iimas.unam.mx (M.E.-A.); sergio@astro.unam.mx (S.M.)}

\firstnote{All authors contributed equally to this work.}

\abstract{
We propose several hierarchical graphs that represent the semantic
relations between physical theories, their fundamental constants and units
of measurement. We begin with an alternative representation of Zel'manov's cube of fundamental constants as a concept lattice. We then propose the inclusion of a new fundamental constant, Milgrom´s critical acceleration, and discuss the implications of such analysis. We then look for the same fundamental constants in a graph that relates magnitudes and units of measurement in the International System of Units. 
This exercise shows the potential of visualizing hierarchical networks as a tool to better comprehend the symmetries, interrelations and dependencies of physical magnitudes, units and theories. New regimes of application may be deduced, as well as an interesting reflection on our ontologies and corresponding theoretical objects.
}


\begin{document}

\section{Introduction}
\label{introduction}

The first study dedicated to the natural units system made up of the three fundamental constants, named $c$, $G$ and $\hbar$, was written as a humorous present to a female student that young George Gamow, Dmitrii Ivanenko and Lev Landau courted back in 1927 \citep{okun2002article, BronsteinSovietPhysics}. They hypothesized what would happen if any of these values were different from what we measure \citep{gamow2002world}. George Johnstone Stoney in 1874 and Max Planck in 1899 had already proposed defining metric standards in terms of fundamental constants: a basic system of units for distance,
time, mass and temperature from which all the other quantities---based on the constants $c,G,h,k$ or $e$---can be normalized to $1$. 
Some words of Max Planck are as follows~\citep{barrow2002constants}:

\begin{quote}
``These quantities retain their natural significance as long
as the law of gravitation and that of the propagation of light in
a vacuum and the two principles of thermodynamics remain valid; they
therefore must be found always to be the same, when measured by the
most widely differing intelligences according to the most widely differing
methods''.
\end{quote}

\begin{figure}[H]

\includegraphics[width=0.6\textwidth]{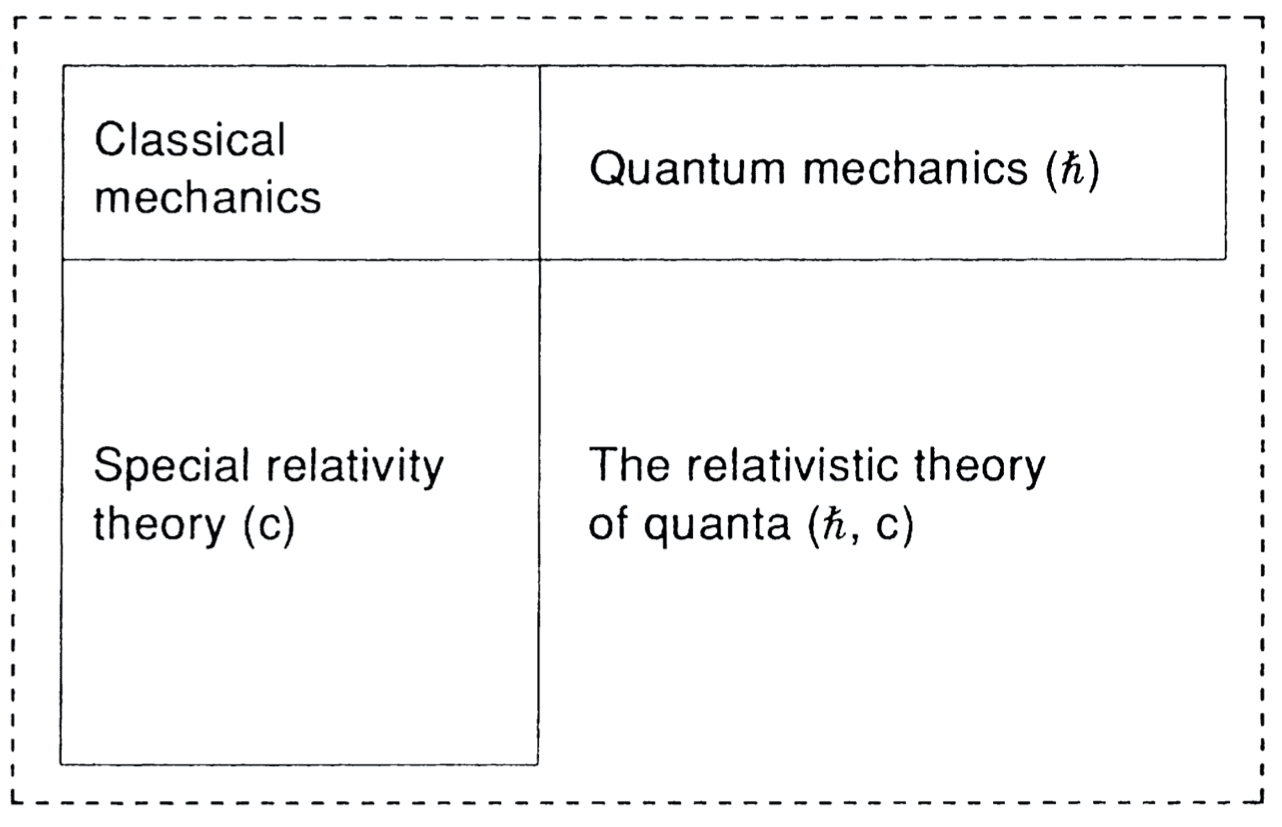}
\caption{Bronstein's space of physical theories.  The image was taken from \citep{BronsteinSovietPhysics}, under a Creative Commons Copyright License.}
\label{Bronstein01}
\end{figure}

\begin{figure}[H]

\includegraphics[width=0.6\textwidth]{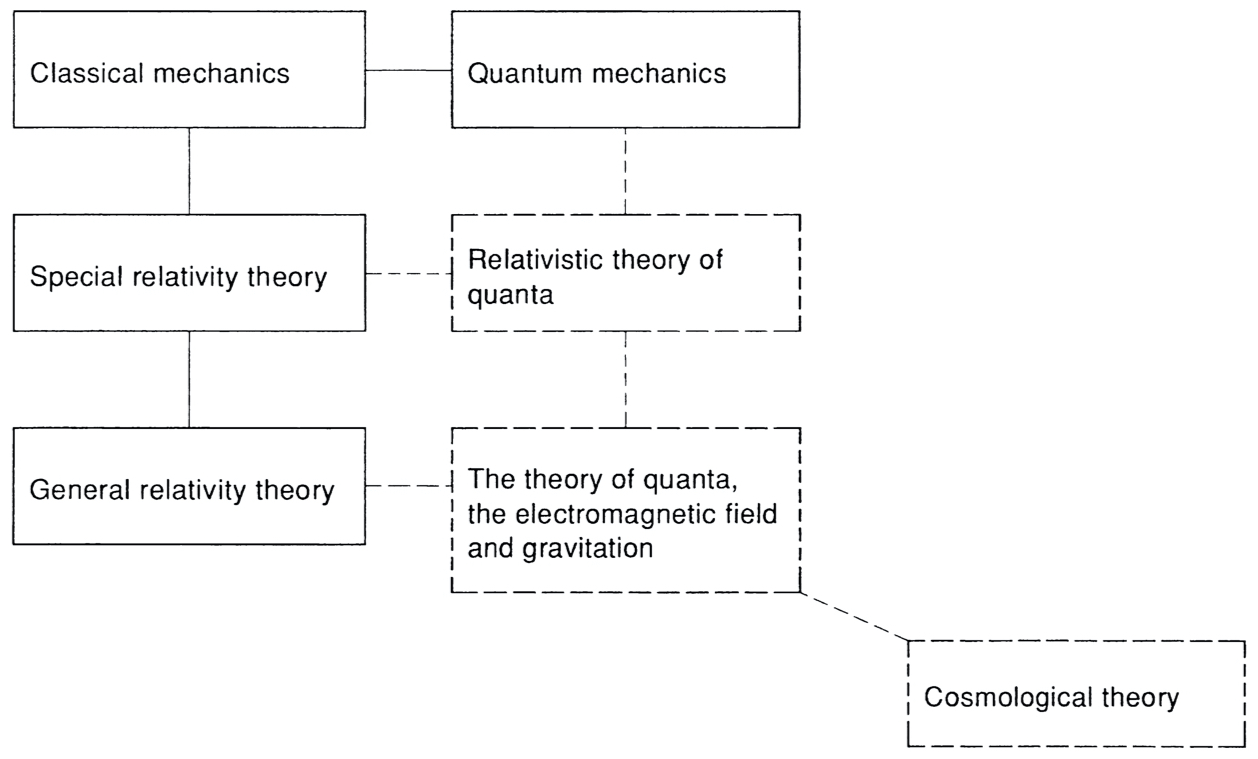}

\caption{Bronstein's space of physical theories and their correlation to cosmology. ``Continues lines correspond to already existing theories. Dotted lines correspond to still unresolved problems'' (Image taken from~\citep{BronsteinSovietPhysics}~under a Creative Commons Copyright License).} 
\label{Bronstein02}
\end{figure}

Those ideas may have found no practical use or meaning and attracted little
attention at the time, but were shared by M. Bronstein, a well respected
mate of the {{Jazz Band} 
} (Landau, Ivanenko and Gamow), who had talked about the limits of theories and was deeply concerned by the general physical picture of the world as a whole. He was one of the few that continued research on gravitation during the 1930s, talking about a unified theory that would marry the field theory of quanta, the electromagnetic field and gravitation. 
This unified theory would be a cosmological theory. Einstein too was looking for a unified theory, but by trying to deduce quanta from his generalized theory of relativity, while Bronstein recognized both theories as equally fundamental that needed some kind of synthesis \citep{BronsteinSovietPhysics}. In  \citep{1933UsAsT...3....3B}, he published two diagrams, reproduced in Figures~\ref{Bronstein01} and~\ref{Bronstein02}, where he maps the world of physics. 
In Figure~\ref{Bronstein01}, the areas of applicability of classical mechanics, quantum mechanics and special relativity are delimited by intersecting rectangles. The dotted lines correspond to a broader area for the not yet well-defined quantum relativistic theory that would contain the other three areas. Such theories are characterized by the presence of the two fundamental constants $c$ and $\hbar$. Figure~\ref{Bronstein02} is not about areas of applicability but relations between the theories themselves, including general relativity theory and the theory of quanta-electromagnetic field-gravitation, which would in turn be the ``cosmological theory''. The dotted lines correspond to unsolved problems. Notice that this representation does not include Newton's gravitation, nor a path from $G$-theory to $cG$-theory.

\vspace{-12pt}

\begin{figure}[H]

\includegraphics[width=0.6\textwidth]{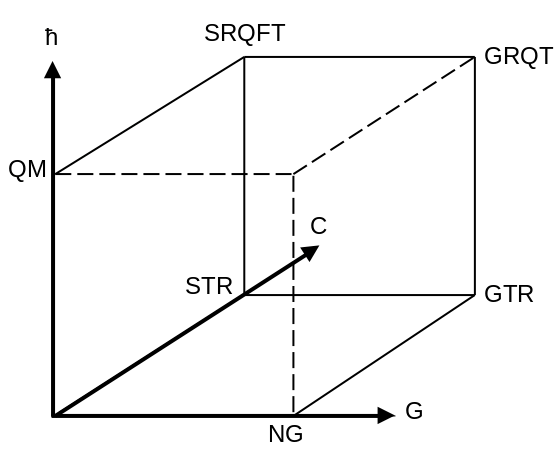}

\caption{{Redrawing} 
 of Zel'manov's cube of physical theories in the cG$\hbar$ coordinate system, as described by~\citet{BronsteinRussian}. Acronyms are as follows: NG: Newton's theory of gravity;
STR: Special theory of relativity; QM: Quantum mechanics; GTR: General theory of relativity; SRQFT: Specially relativistic quantum field theory, GRQT: General relativistic quantum theory. Notice there are two corners without a label.}

\label{Okun}
\end{figure}

\textls[-15]{{According to \cite{BronsteinSovietPhysics}, to represent Bronstein's ideas in three dimensions, in 1967 A.L.
Zel'manov~\cite{ZelmanovKosmologiya} drew} the cube of physical theories
(Figure~\ref{Okun}), which was reproduced
by \citet{okun1991fundamental} and \mbox{\citet{barrow2002constants}}. This
cube consists of three orthogonal axes $(G,1/c,\hbar)$. The origin
(0,0,0), corresponds to the limits $G\rightarrow0$, $1/c\rightarrow0$ and $\hbar\rightarrow0$, representing classical particle mechanics in the absence of extreme fields. In other words, the speed
of light $c \rightarrow \infty$---or more formally the speed of propagation
of interactions---is infinite, and Planck's constant does not
describe the nature of classical physics, so $\hbar \rightarrow 0 $ and, given
that there is no gravitational description of nature, $ G \rightarrow 0 $.
Note that the fundamental constants of physics $ c, \ \hbar $ and  $G $
are independent of each other. This principle about fundamental constants then implies
that there is no mathematical formula in which one of them can be obtained by
the other two. }

\begin{figure}[H]

\includegraphics[width=0.8\textwidth]{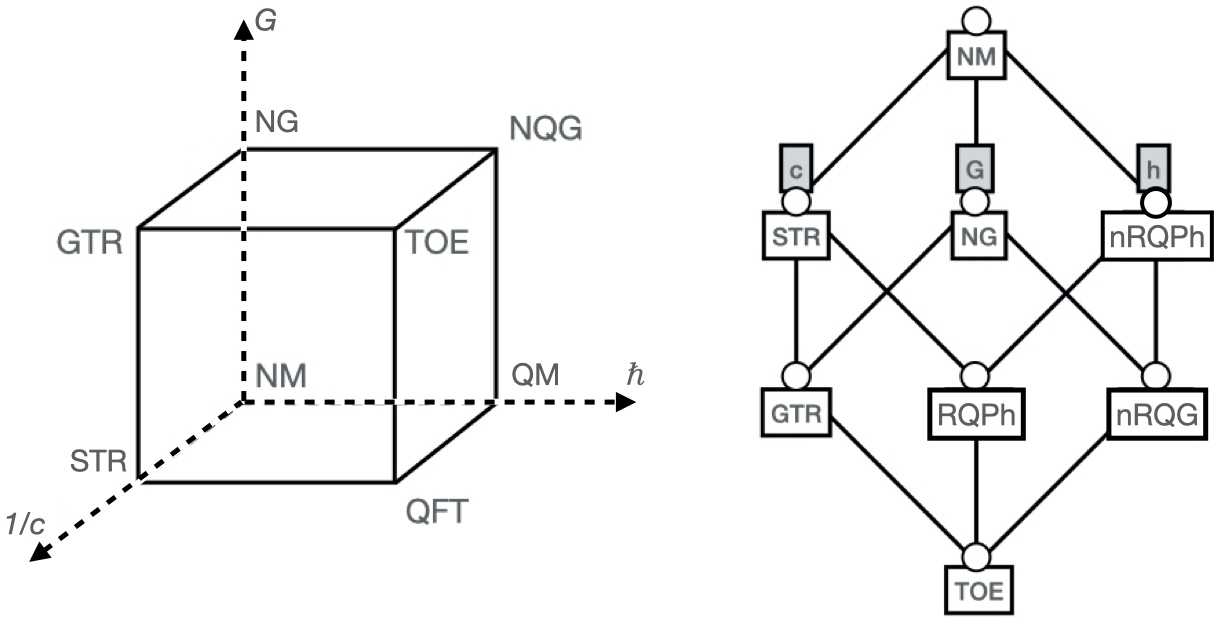}

\caption{{{On the left:} 
the cube of physical theories by Okun \citep{okun1991fundamental}, where
TOE, NM and QFT appear, and the c axis is now the 1/c axis.} The
abbreviations in the diagram stand for NM{---}
Newtonian Mechanics; NG---Newtonian Gravity; STR---Special Theory of Relativity;
QM---Quantum Mechanics; GTR---General Theory of Relativity; NQG---Non-relativistic Quantum Gravity; QFT---Quantum Field Theory; TOE---Theory
of Everything. The figure on the right corresponds to the lattice of physical theories. Unlike a cube with
coordinate axes, where Newtonian Mechanics (NM) lies at \guillemotleft the
origin\guillemotright{} (0,0,0), here it lies in the supreme node
while the Theory of Everything (TOE) is at the bottom node. We have generalized even more the names for NQG to non-Relativist Quantum Gravitation (nRQG), QM to non-Relativistic Quantum Physics, and QFT to Relativist Quantum Physics (RQPh) in order to accommodate other models that would describe such regimes.}
\label{CuboReticulo}
\end{figure}

{Figure~\ref{Okun} is a translated copy from the original diagram in Russian of Zel'manov's cube (also called Okún's cube, ``physical theories cube'' or ``hypercube of physical units''). 
There are two empty vertices, no Newtonian Mechanics, no TOE and no QFT. Instead, there is a General Relativistic Quantum Theory. This situation was changed by Okun in the left side of Figure~\ref{CuboReticulo}.} Newtonian gravitation is then located at the vertex $(G,0,0)$,
while special relativity is at $(0,1/c,0)$ and general relativity corresponds to  $(G,1/c,0)$. Non-relativistic quantum mechanics
is located at $(0,0,\hbar)$, quantum field theory at $(0,1/c,\hbar)$
and the Theory of Everything (TOE) is located at the corresponding corner
$(G,1/c,\hbar)$ {(see also}~
\citep{Duff:2001ba}). This representation by means of axes can take
the reader to interpret a smooth and continuous gradient in the values of the constants, which
is incorrect because the values of each fundamental constant of physics is unique and,
in principle, they do not have a spatial or temporal change.
{{However}
, it should be noted
that there are proposals for temporal variations in the gravitational constant \citep{PhysRev124925}, in the speed of light \citep{magueijo2003faster} and even in the fine structure constant {cf.}~\citep{barrow2002constants,martinez-mendoza}.}

{The interest in the relations between fundamental constants was also expressed in the western world by Dirac \cite{Dirac:1937ti,Dirac1938,2007arXiv0705.1836R}, who proposed his large number hypothesis, observing relations between ratios of fundamental constants and size scales of the Universe, such as the gravitational constant being inversely proportional to the age of the universe $G \propto 1/t$, suggesting THAT constants could be subjected to the proportions of the Universe's scales.}

{The Boltzmann constant is not considered
fundamental since it serves only to transform temperature to energy values. However, {Huggett et al}. \citep{huggett2020beyond} proposed a hypercube with a fourth direction, labeled \textit{N} for the number of degrees of freedom of a system, but related to Boltzmann's constant as a way of linking information theory and its entropy.}

As we will show, lattices are an alternative and more powerful multidimensional representation. Physical theories can be classified on the basis of many different approaches such as their axioms, their models or their parameters, fundamental constants and symmetries  using different formalisms such as set theory or category theory. 
These classifications can also be represented using diagrams. Recent work
by \citet{espinosa2022visualizaciones, EspinosaGOL2023} show the benefits
of using concept lattices and computing technologies to create complex
hierarchical networks that represent relations between models of physical
theories. ({See also} \citet{ MendezDominios2012} for a similar study on
genetics.)  In this work, more than a dozen theoretical grids were
generated that relate more than 44 gravitational and space-time theories.
({These interactive} networks can be found on{ the following web page:} 
 \url{http://remo.cua.uam.mx /vis/Exploratorium}, accessed on 1 July 2024).
 These conceptual maps allow students to locate their working models in a
 sea of concepts; identify the foundations of theoretical models and
 theoretical branches; and differentiate classes of theories, classes of
 concepts as well as sub-specialties. By studying these networks, it was
 possible to refine our definition of theoretical specialization and the
 theorizing relationship, as well as to identify a certain parsimony in
 scientific progress. These networks, called lattices, are obtained using
 {{Formal Concept Analysis} 
} (FCA), a mathematical theory based on  Garrett Birckoff’s lattice theory and Galois' connection that relates concepts within a formal context, {see}~\citep{FCAGanterWille1999,ganter2005formal}. 

\vspace{-6pt}

In the present article, we focus on the presence of fundamental constants in law statements of the most relevant physical theories such as quantum theory, gravitation
and relativity. Under the structuralist view of the philosophy of science, these constants are considered distinguished elements. These are variables that always take an exact value, under certain conditions,
either as quantities of some physical variable or product numbers
of ratios between quantities.
Dimensionless constants, like the
fine-structure constant $\alpha\approx 1/137$ that quantifies
the strength of electromagnetic interactions between charged particles,
take the same value regardless of the system of units, and
in that sense they are universal. Instead, the dimensional constants
refer to properties of classes of physical systems whose value, which is constant in time, depends on the system of selected units.
As we know, the value of the gravitational constant in the SI is {$G$ =
6.67384 $\times$ 10$^{-11}$~\mbox{N $\cdot$ m$^{2}$/kg$^{2}$}}. 
In astrophysics,
where distances are measured in parsecs $(pc)$, velocities
in {km/s$^{2}$} and the masses in solar units $M_{\odot}$, the value
is $G$ = 4.3009 $\times$ 10$^{-3}$~$pc/M_{\odot}$(km/s$^{2}$), while in the system
of natural Planck units, the constants are normalized,
where $G=1$.

Not all referential constants are universal, and
there are constants specific to particular systems; the speed of light depends on the medium and the speed
of sound even more. The specific heat is proper to the material, and charge acts only upon specific particles. {(See~\cite{universe6100166,tonti2013mathematical} for other discussions on the nature of fundamental constants and their graphic representations.}) Constants
are characteristic of certain regimes. The extension of a particular regime
tells us how more fundamental or less fundamental a constant is. For example, the charge of the electron $e$ belongs to micro physics.
Others, like Boltzmann's constant, belong to macro physics,
and others like $G$ and $c$ are assumed to be independent of the
scale.  G is a universal constant since it does not depend on the medium, the material or the scale.
Under some theories like Modified Newtonian Dynamics (MoND), Milgrom's acceleration is a fundamental constant \citep{2011MPLA...26.2677B} that defines the low-acceleration regime \citep{2012Entrp..14..848H}.
A first study on the incorporation of Milgrom's acceleration as a fundamental constant in a diagram that semantically relates it to the other three fundamental constants, $c$, $h$ and $G$, can already be found in \cite{espinosa2022visualizaciones}.

The article is organized as follows. Section~\ref{formal} explains in a
nutshell the mathematical theory called Formal Concept Analysis (FCA),
which helps us relate concepts hierarchically and present them as lattices,
showing how it can be applied to Zel'manov's cube. Section~\ref{structuralism} continues to pinpoint some relevant aspects of structuralism, which is our theoretical framework and comes in handy for our analysis, mostly presented in Section~\ref{physical-theories}. Section~\ref{international-system} takes a small detour around the international system of units and their relation to fundamental constants in order to present an alternative representation using FCA. We then take the dual graph to include some 47 derived magnitudes as objects. In this multidimensional space, we can also locate the fundamental constants as distinguished objects and evaluate their semantic hierarchical relations. Finally, we share our insights and possible follow-ups in Section~\ref{insights}.

\section{Formal Concept Analysis}
\label{formal}

Lattices are tree-like, hierarchical networks that have the peculiarity that for any pair of nodes there is one and only one supreme and infimum node. Conceptual lattices are hierarchical networks whose nodes represent formal concepts. The meaning of a formal concept $C: <B,A>$ is given by a subset of certain attributes $A\subseteq M$ called the intention of the concept, and by all those objects $B\subseteq G$ that have those attributes, the extension of the concept. Formal concepts are dependent on the context in which they are found. A formal context $(G,M,I)$ is a binary table with an incidence relation $I\subseteq G\times M$, where a set of objects $G$ is related to a set of attributes $M$. Formal Concept Analysis (FCA) is a mathematical theory aimed at data analysis and classification, developed by Rudolf Wille and his group since the 1980s at the Technical University of Darmstadt, Germany. It is a theory derived from the lattice theory of \citet{birkhoff1940lattice} incorporating the monotone Galois connection
\citep{FCAGanterWille1999} in order to obtain formal concepts and their hierarchical relationships, levels, sub-concepts and supra-concepts; assign labels; explore objects and attributes; obtain implications and associations; calculate meanings; and obtain indexes of stability, conceptual probability, etc. 
FCA has developed many techniques and applications that nowadays are widely used in knowledge representation and reasoning, information retrieval, concept classifiers, knowledge processing, knowledge discovery, ontology engineering, pattern structures recognition, machine learning, science communication and in the analysis of theoretical structures and semantic networks \citep[cf.][]{ganter2005formal,2016arXiv161102646K,KUZNETSOV2018202,ganter2002methods,Priss2007,ganter2016conceptual,KumarFCASurvey2016}.

Two derivation operators represented by $\ll$ $'$ $\gg$ have the function of assigning sets of objects to sets of attributes or sets of attributes to sets of objects. If a set of objects $B=(g_{1},\ldots g_{n})$ has (shares)
certain attributes $B'=(m_{1},m_{2},\ldots m_{m})$,
and the corresponding objects to such attributes---say $B''$---
are exactly $B$, then $\left\langle B'',B'\right\rangle $
is a formal concept from the context.
In a dual way,
let $A$ be a set of attributes and $A'$ the set of all objects that share such attributes. If after applying again the operator $'$ to $A'$ we obtain $A=A''$, then $\left\langle A',A''\right\rangle$
is a {{formal concept}} of the context. This double operation $''$  
is known as {{derivation}}.  
So, within a formal context we can find formal concepts $\left\langle B_{1},A_{1}\right\rangle$, $\left\langle B_{2},A_{2}\right\rangle$,
$\left\langle B_{i},A_{i}\right\rangle$, within all the pairs of sets from the context $(G,M,I)$.

Formal concepts are then ordered by comparing their extensions and intentions:
\[
\left(A_{i}\text{\ensuremath{\subseteq}}A_{j}\right)\text{\ensuremath{\Longleftrightarrow}}\left(\left\langle B_{i},A_{i}\right\rangle \geq\left\langle B_{j},A_{j}\right\rangle \right)\Longleftrightarrow\left(B_{i}\supseteq B_{j}\right)
\]

This is a partial order relation with the properties of reflexivity, antisymmetry and transitivity. We can draw this hierarchical relation between the concepts as links between nodes, assigning in this manner a level to each concept starting from the bottom node, which is always unique. For any lattice, any pair of nodes has a unique common supreme node and a unique common infimum node. Thus, the top concept contains, if any, the attributes shared by all objects, while the bottom concept contains all the attributes and generally no objects. 

{The analysis of the formal context and lattice visualization is performed
using Concept Explorer (\textit{{ConExp 1.3}} available on
\url{https://conexp.sourceforge.net/}, accessed on 15 March 2024), an
easy-to-use \textit{{Java}
} native application that also allows exploring different dependencies that exist between attributes \cite{ConceptExplorer}. A table is offered to be filled as a formal context, naming objects and attributes, and an interactive lattice is built upon the calculation of concepts.}

\section{Structuralism}
\label{structuralism}

The structuralist view of philosophy of science follows a set--model--theoretic approach to analyze theories, using three units of analysis: theoretical models, theoretic elements and theoretical networks \citep{ExploracionesMoulines1982,balzer2000structuralist,Balzer2012,abreu2013bibliography}. Theoretical models are concepts too, with intentions and extensions. Intentions are formed by n-tuples of classified attributes that comply with certain axioms of the theory, while extensions are exemplary physical models that represent them as objects. 
The attributes include domains, relations between domains, constraints and their interpretations under a context. Classes of models such as previous models, potential models, actual models and specialized models can be identified in a lattice configuration. In physics, previous models may be models of space-time theories, protophysics, topology, geometry, etc. (e.g., \cite{bunge1967foundations}, on the notion of protophysics). Potential models include new physical domains such as mass or electric charge, as well as certain distinguished elements. Actual models enter the physical laws that relate the given domains, while specialized models introduce constraints, conditions and other peculiarities of a given scenario. 
It is in the actual models where the laws of movement and field equations are found. The fundamental constants are present in such equations, carrying nature's impositions that we measure, observe or hypothesize; as such distinguished elements found in several theories, they can represent families of models that contain them. Law statements may occupy several of these constants; in that sense, they share regimes of applications.

\section{Physical Theories Lattice}
\label{physical-theories}

Following the idea that distinguished elements such as fundamental constants may represent classes of potential models of physical theories, and using FCA to classify theories and construct theoretical networks, we proceed to interpret Zel’manov's cube as a concept lattice 
(Figure~\ref{CuboReticulo}). 
In the right side representation of Figure~\ref{CuboReticulo}, the positions of the nodes are reflected on the horizontal plane: Newtonian Mechanics (NM) is no longer at the origin but at the top node, while the Theory Of Everything (TOE) has moved to the bottom node. 
{We have called ``non-Relativistic Quantum Physics'' 
(nRQPh) and ``Relativistic Quantum Physics'' (RQPh) to all classes of physical theories that require \( \hbar \) and \( c \) in 
their mathematical description.  As such, QM and non-relativistic QFT enter a general classification scheme called nrQPh and QFT enters RQPh. This allows the inclusion of other possible theories that are not QM or QFT into the proper class.}

\begin{figure}[H]

\includegraphics[width=0.6\textwidth]{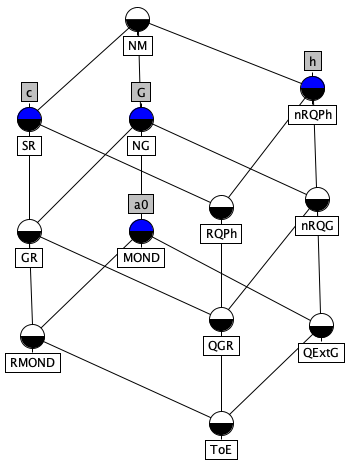}
\caption{Lattice of fundamental constants includes Milgrom's constant
$a_{0}$ and four new model theories: Extended Gravitation (ExG), Extended Relativistic Gravitation (ExRG) and Extended Quantum Gravitation (ExQG). Relativistic Quantum Gravity (QGR)
replaces the original Theory Of Everything (TOE), leaving the new TOE in
the lowest node, which does not exist either, but would have to include
all fundamental constants---that is, all regimes. Blue half nodes indicate a new attribute; black half nodes indicate a new object.}\label{ConstantesGood}
\end{figure}

Figure~\ref{ConstantesGood}
incorporates Milgrom's acceleration $a_{0}\approx1.2\times10^{-10}$m/s$^{2}$ as a fundamental constant \citep{2011MPLA...26.2677B}, generating four additional nodes that suggest the possible existence of new application regimes. These regimes can be delimited using phase state diagrams or conceptual spaces.
Over the past 40 years, dozens of alternative theories of gravitation that
consider Milgrom's acceleration $a_0$ as a relevant parameter have been
proposed, both relativistic and non-relativistic. Researchers {such as}
 \citet{1983ApJ...270..365M,Bekenstein1984ApJ,2011MNRAS.411..226M,MONDReview,PhysRevLett.127.161302} have tested their extended theories across all astrophysical scales with promising results and predictions.

The introduction of Milgrom's acceleration 
constant \( a_0 \) as a fundamental constant 
in physical theories has been considered since 
it was first introduced as a modification or extension of Newtonian gravitation to understand different astrophysical phenomena instead of using non-baryonic dark matter. {See, e.g.,}~\citep{ 1983ApJ...270..365M,Milgrom2,Bekenstein1984ApJ}.  As explained by \citet{2011MPLA...26.2677B}, the constant \( a_0 \) has a strong fundamental nature in exactly the same form as the physical constants \( G, c \) and \( \hbar \).  These authors also showed that the very well-known relation between \( a_0 \) and the Hubble constant \( H_0 \) at the present epoch  and the cosmological constant \( \Lambda \) given by \( a_0 \approx c H_0 \approx c \sqrt{\Lambda} \) \, \citep{1983ApJ...270..365M,pedago}  is a coincidental relation and only occurring at the present epoch in the evolution of the Universe.  This fundamental fact was also found in relativistic extensions of MOND, as described by \citet{barrientosbernalmendoza,barrientosmendozapadilla}.  In other words, to account for different astrophysical and cosmological phenomenology without postulating the existence of dark matter and/or energy by using an extended gravitational theory, the introduction of the constant \( a_0 \) is required to understand a new regime of gravitation at low acceleration scales \( \lesssim a_0 \).

\begin{figure}[H]

\includegraphics[width=1\textwidth]{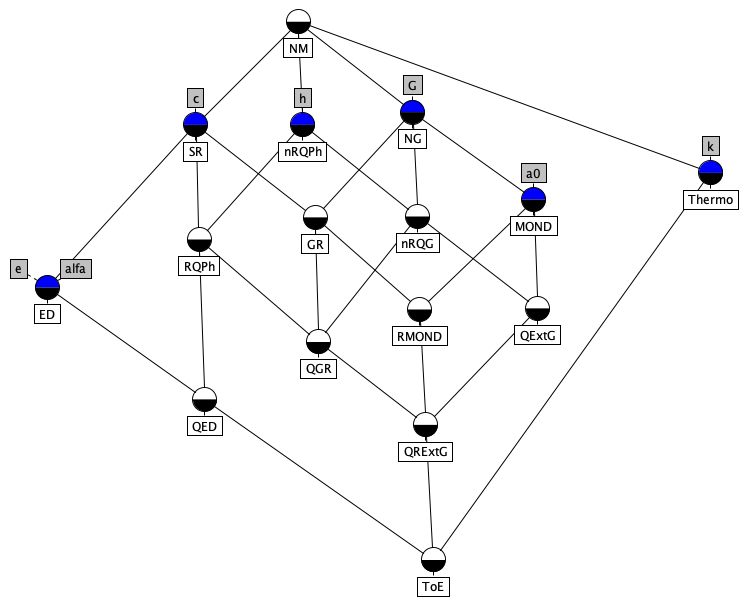}

\caption{Concept lattice with more constants: the electric charge $e$, Boltzmann's constant $k$ and Milgrom's acceleration $a_{0}$ call for the Electrodynamic Theory, Quantum Relativistic Extended Gravitation and Thermodynamics.}
\label{ReticuloConstantes}
\end{figure}

We can also envision theories that would attempt to describe microscopic interactions in a low-acceleration regime or even a relativistic quantum low-acceleration regime. This would represent a new TOE, requiring the consideration of all relevant attributes. Adding another constant to the set of distinguished elements may counter some expectations of reducing the number of fundamental constants needed to achieve a TOE or reducing them to dimensionless constants, as was Einstein's intention during his later research period.
Nevertheless, exploring the consequences of such addition, including constants like $e$, $k$ or $\mu_{0}$ as shown in Figure \ref{ReticuloConstantes}, is a worthwhile endeavor.

\section{The International System of Units}
\label{international-system}

According to the International System (SI) of units, there are seven
 basic units of measurement: kilogram (kg), second (s), meter (m),
ampere (A), Kelvin degree (K), candela (cd) and mole (mol). These units
are defined from our ability to count and from combinations and ratios of measured quantities of length (L), time (t), mass (M), electric current (I), luminous intensity (J) and temperature (T). The charge of the electron $e$ is measured in Coulombs or Ampere per second: ($[e]=[C]=[As]$);
Avogadro's number $N_{A}$ is measured in moles: ($[N_{A}]=[mol^{-1}]$). {One mole of a substance} is the amount of it that contains a number
specific to elemental entities. Although it is an amount of substance,
we cannot express it as a mass (for example, grams) until
the elementary entity is specified. Those entities can be either atoms, molecules,
ions, electrons, other particles or specific groups of such
particles  \citep{resnick1974fisica}. The proportionality factor between the kinetic energy of a gas
and its temperature is Boltzmann's constant $k_{B}$, where ($[k_{B}]=\left[ML^{2}/Tt^{2}\right]$).
For Planck's constant \textrm{$h$}, the units are ($[h]=[ML^{2}/t]$); for
the speed of light \textrm{$c$}, ($[c]=[L/t]$); while the oscillation frequency of a {Caesium} atom $\Delta\nu_{Cs}$ in its ground state defines the second ($[\Delta\nu_{Cs}]=[s^{-1}]$).

Other unit systems do not include $k_{B}$ or $N_{A}$ between
their axiomatic components, since it is considered that they are derivable from the
base set and do not describe properties of the universe but factors
of proportionality.

\newpage

\begin{figure}[H]

\includegraphics[width=0.8\textwidth]{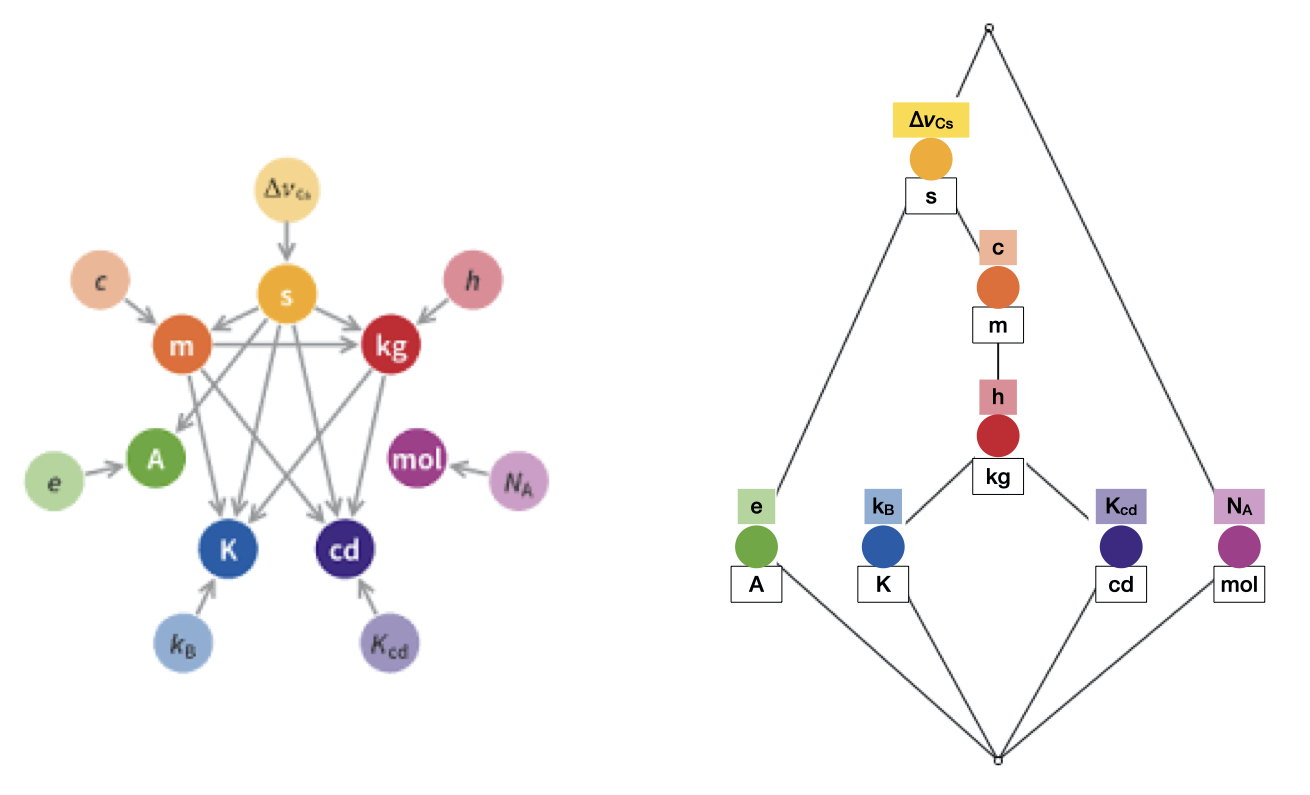}

\caption{Two representations of the relationships between units of the international system through fundamental constants. The left image was taken from \citet{PisantyNewSI} under a Creative Commons Copyright License and the right one is our proposal. The colors used in the
lattice follow Pisanty's selection to facilitate comparison
between both diagrams. }
\label{IS}
\end{figure}

The SI went through a revision during the second decade of the 21st century
by the International Committee for Weights and Measures, achieving that
all units, including the kilogram, will be defined from fundamental constants that can be obtained by experiment and not from exemplary physical objects. In Figure \ref{IS},
two representations of the relationships between the units are shown for the SI.
The figure on the left side, by \citet{PisantyNewSI}, and published
by \citet{lockwood2016quantifying}, presents a circular or heptagon configuration with some disconnected nodes. The figure on the right side is an alternative representation obtained with FCA, which emphasizes the hierarchical level of the constants
as attributes and units as objects. The lattice was constructed using the same relations indicated in Pisanty's diagram and simply adding a top and bottom nodes. It is presented with the same colors for easy comparison. In the lattice, the attributes are the fundamental constants and the basic units are objects. The resulting concepts are as follows: Time, Space, Mass, Temperature, Electric Intensity, Amount of substance and Luminous Intensity. 
At this point, we draw our attention to the absence of $G$ in the definition of the basic units.

\begin{figure}[H]

\includegraphics[width=0.7\textwidth]{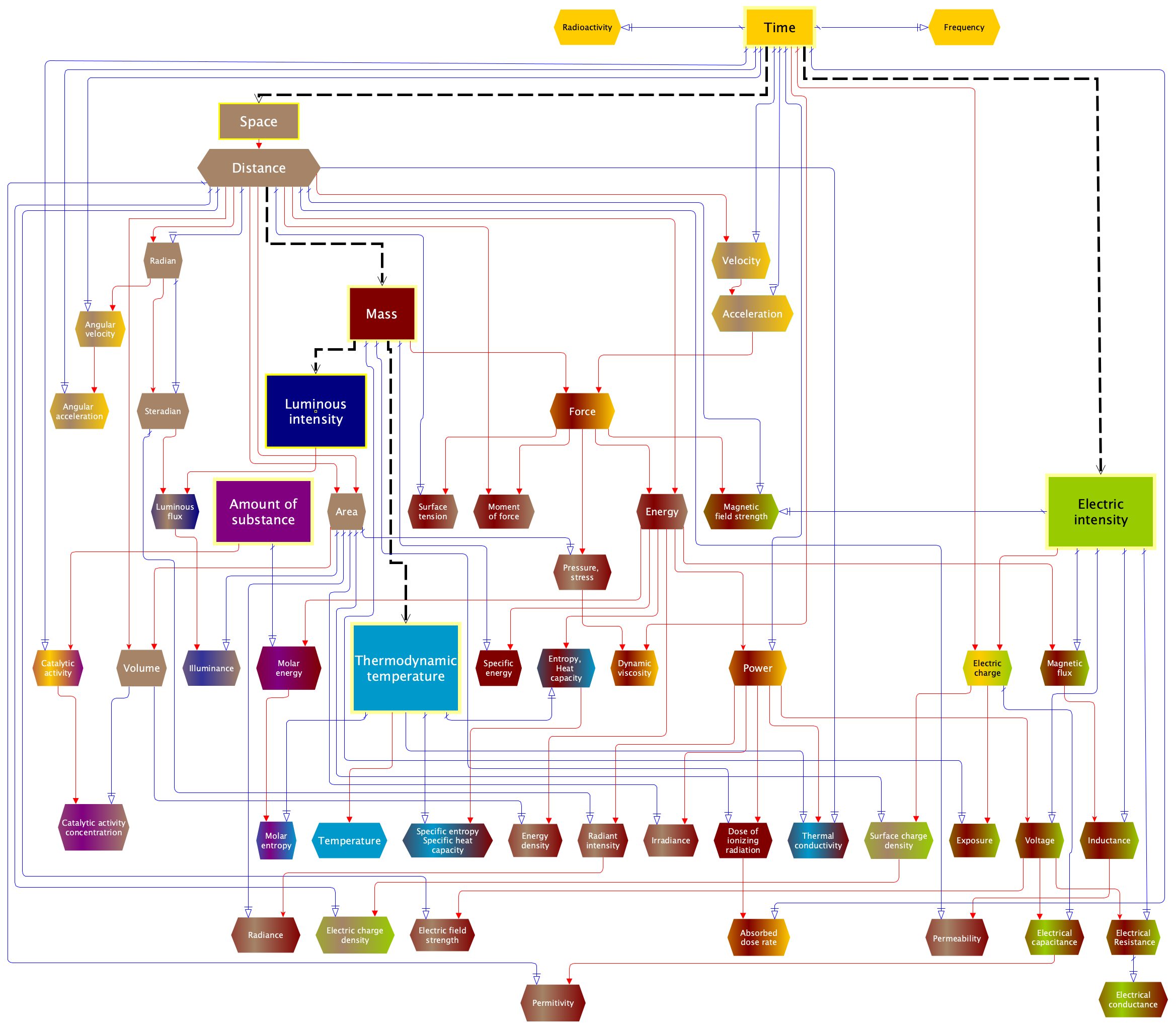}

\caption{{Hierarchical} 
 network for derived magnitudes. Red arrows represent multiplication of basic units. Blue arrows represent a division of basic units. Dotted bold arrows keep the hierarchical relations brought from Pisanty's network.}
\label{DerivadasColores}
\end{figure}

In this diagram, our ontology is based on phenomenology. Our units are artificial objects that we create from the constancy of our observations. 
 We can now build our world based on those objects. In Figure \ref{DerivadasColores}, we built a network taking basic units as attributes and magnitudes as objects---relating derived magnitudes such as force, energy, power---and represent their hierarchical placement using two different kinds of relation: multiplication (red arrows) and division (blue arrows). Figure \ref{DerivadasColores} also gives nodes a combination of colors according to their parents' color.
There is still a hierarchy and top-down directionality, but different arrows represent different relations between units. We can find out the basic units of any magnitude just by following the arrows backwards, adding the red ones and subtracting the blue ones.

\begin{figure}[H]

\includegraphics[width=0.62\textwidth]{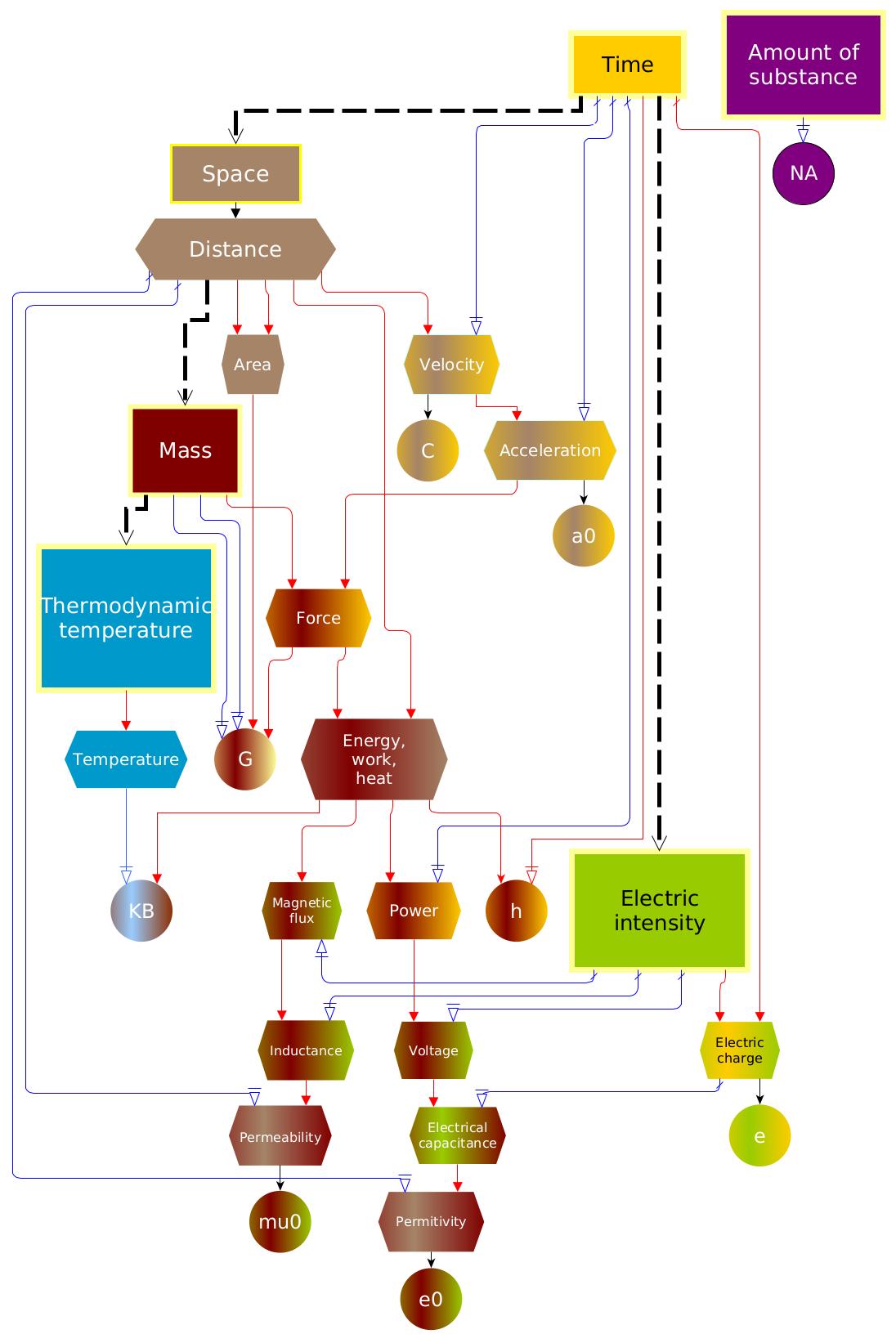}

\caption{{Hierarchical} network for derived magnitudes with fundamental constants (circle nodes).}
\label{DerivadasConstantes}
\end{figure}

In this new space, we can also locate the fundamental constants such as $G$, $c$, $h$, $k$ and $a_{0}$. Figure \ref{DerivadasConstantes} shows a clarified network where only attributes relevant to the fundamental constants are shown. Here, constants are derived from basic units, showing how some of them do not have a magnitude properly associated.
\vspace{-12pt}

\section{Insights and Follow Ups}
\label{insights}

This paper presents a novel hierarchical multidimensional representation of
physical theories, fundamental constants and units of measurement using
Formal {Concept} Analysis (FCA). By re-examining Zel'manov's cube and introducing Milgrom's critical acceleration as a new fundamental constant, we aim to provide a more nuanced understanding of theoretical structures and their interrelations.

{Graphic representations of relations between concepts are surely powerful aids to a better communication of ideas, as well as aids in the scientific thinking process and search for symmetries. Here, we define a formal context and identify formal concepts, which are not random combinations of objects and attributes but intentions and extensions. Researchers in physics and metatheoretic studies as well as the public may benefit from this new algorithmic technique of FCA, which increases our confidence on a certain network of concepts: a concept lattice. This choice of methodology to represent multidimensional inter-theoretical relations, in a historical discussion that clearly has not been solved, may give a refreshing perspective to the problem of classifying theories.}

We have seen that such networks suggest the possible existence of other theories that would be applied to new regimes (i.e., Mondian quantum theories). Drawing phase-space diagrams of such regimes and finding their concept spaces would be a suitable follow-up for this research. 

{Certainly, the construction of formal contexts depends on the selection of our premises as fundamental, as well as the reconstruction method, which can be subtractive or additive. Algebraic formulations with global variables or differential formulations also exhibit different aspects of the theoretical structure. Notations may also take a part of our representations. This is why we have to state which approach we are taking: the semantic-set-model-theoretic approach of the meta structuralist view developed in \cite{Balzer2012}}. 

Focusing on a distinguished element allows us to reduced the kind of attribute to the minimum, leaving out domains, fields, special conditions and equations, etc. Fundamental constants as representatives of potential models embrace a huge amount of actual and specialized classes of models. We are here presenting just the most general classes of models and some or their relations that contain our to-date physics.

There is an evident decision on the selection of a proper ontology when we decide whether basic units or basic constants are to be our attributes and the abstractions (objects) that can be produced with them. It is evident that the use of artificial units allows a simple derivation method and layout of relations between magnitudes. 

Future research could focus on integrating other fundamental constants and refining the lattice structure to accommodate more specialized models. Developing computational tools to automate the generation and analysis of these lattices will be crucial in enhancing their practical utility and accessibility.

In such a direction, we can further analyze in the search for possible loops, application regimes and visual aid in theoretical studies. In our way, we have valued the interplay that the relation between quantities and units have and also considered the space where fundamental constants live in relation to basic and derived quantities.

\vspace{6pt}
\authorcontributions{{Conceptualization, M.E.-A.; Methodology, M.E.-A.  and S.M.; Validation, M.E.-A. and S.M.; Formal analysis, M.E.-A. and S.M.; Investigation, M.E.-A. and S.M.; Resources, M.E.-A. and S.M.; Writing – original draft, M.E.-A. and S.M.; Writing – review and editing, M.E.-A. and S.M.; Visualization, M.E.-A.; Funding acquisition, M.E.-A. and S.M. All authors have read and agreed to the published version of the manuscript.} 
}

\funding{{This research was funded by CONAHCyT Estancias Posdoctorales por México Convocatoria 2023(1) grant number 5991306, CONAHCyT grant number 232297, 26344 and PAPIIT DGAPA-UNAM grant number IN110522.} 


}

\dataavailability{{Data are contained within the article.} 
}


\conflictsofinterest{{The authors declare no conflicts of interest.} 
} 
\begin{adjustwidth}{-\extralength}{0cm}

\reftitle{References}

\PublishersNote{}
\end{adjustwidth}

\end{document}